\documentclass[sigconf]{acmart}
 \usepackage{amsmath,amssymb,amsfonts,bm}
\usepackage{graphicx}
\usepackage{textcomp}
\usepackage{xcolor}
\graphicspath{ {pic/} }
\usepackage{float}
\usepackage{array}
\usepackage{tabularx}
\usepackage{array}
\usepackage{balance}
\usepackage{marvosym}
\usepackage{multirow}
\usepackage{array}
\usepackage{enumitem}
\usepackage{booktabs}
\usepackage{algpseudocode}
\usepackage[ruled]{algorithm2e} 
\usepackage{xcolor}
\usepackage{pifont}
\usepackage{url}

\AtBeginDocument{%
  \providecommand\BibTeX{{%
    \normalfont B\kern-0.5em{\scshape i\kern-0.25em b}\kern-0.8em\TeX}}}

\setcopyright{acmcopyright}
\copyrightyear{2021} 
\acmYear{2021} 
\setcopyright{acmlicensed}\acmConference[KDD '21]{Proceedings of the 27th ACM SIGKDD Conference on Knowledge Discovery and Data Mining}{August 14--18, 2021}{Virtual Event, Singapore}
\acmBooktitle{Proceedings of the 27th ACM SIGKDD Conference on Knowledge Discovery and Data Mining (KDD '21), August 14--18, 2021, Virtual Event, Singapore}
\acmPrice{15.00}
\acmDOI{10.1145/3447548.3467340}
\acmISBN{978-1-4503-8332-5/21/08}

\begin{document}

\title{Socially-Aware Self-Supervised Tri-Training for Recommendation}

\author{Junliang Yu}
\affiliation{%
	\institution{The University of Queensland}	
	\city{Brisbane}
	\country{Australia}}
\email{jl.yu@uq.edu.au}

\author{Hongzhi Yin}
\authornote{Corresponding author and having equal contribution with the first author.}
\affiliation{%
	\institution{The University of Queensland}
	\city{Brisbane}
	\country{Australia}}
\email{h.yin1@uq.edu.au}

\author{Min Gao}
\affiliation{%
	\institution{Chongqing University}
	\city{Chongqing}
\country{China}}
\email{gaomin@cqu.edu.cn}

\author{Xin Xia}
\affiliation{%
	\institution{The University of Queensland}
	\city{Brisbane}
\country{Australia}}
\email{x.xia@uq.edu.au}

\author{Xiangliang Zhang}
\affiliation{%
	\institution{KAUST}	
	\city{Thuwal}
	\country{Saudi Arabia}}
\email{xiangliang.zhang@kaust.edu.sa}

\author{Nguyen Quoc Viet Hung}
\affiliation{%
	\institution{Griffith University}
		\city{Gold Coast}
	\country{Australia}}
\email{quocviethung1@gmail.com}

\fancyhead{}
\begin{abstract}
	Self-supervised learning (SSL), which can automatically generate ground-truth samples from raw data, holds vast potential to improve recommender systems. Most existing SSL-based methods perturb the raw data graph with uniform node/edge dropout to generate new data views and then conduct the self-discrimination based contrastive learning over different views to learn generalizable representations. Under this scheme, only a bijective mapping is built between nodes in two different views, which means that the self-supervision signals from other nodes are being neglected. Due to the widely observed homophily in recommender systems, we argue that the supervisory signals from other nodes are also highly likely to benefit the representation learning for recommendation. To capture these signals, a general socially-aware SSL framework that integrates tri-training is proposed in this paper. Technically, our framework first augments the user data views with the user social information. And then under the regime of tri-training for multi-view encoding, the framework builds three graph encoders (one for recommendation) upon the augmented views and iteratively improves each encoder with self-supervision signals from other users, generated by the other two encoders. Since the tri-training operates on the augmented views of the same data sources for self-supervision signals, we name it self-supervised tri-training. Extensive experiments on multiple real-world datasets consistently validate the effectiveness of the self-supervised tri-training framework for improving recommendation. The code is released at \url{https://github.com/Coder-Yu/QRec}.
\end{abstract}


\maketitle

\section{Introduction}
Self-supervised learning (SSL) \cite{liu2020self}, emerging as a novel learning paradigm that does not require human-annotated labels, recently has received considerable attention in a wide range of fields \cite{oord2018representation,chen2020simple,velickovic2019deep,zhai2019s4l,qiu2020gcc,han2020self,lan2019albert}. As the basic idea of SSL is to learn with the automatically generated supervisory signals from the raw data, which is an antidote to the problem of data sparsity in recommender systems, SSL holds vast potential to improve recommendation quality. The recent progress in self-supervised graph representation learning \cite{velickovic2019deep,you2020graph,jin2020self} has identified an effective training scheme for graph-based tasks. That is, performing stochastic augmentation by perturbing the raw graph with uniform node/edge dropout or random feature shuffling/masking to create supplementary views and then maximizing the agreement between the representations of the same node but learned from different views, which is known as \textit{graph contrastive learning} \cite{you2020graph}. Inspired by its effectiveness, a few studies \cite{zhou2020s,yao2020self,wu2020self,ma2020disentangled} then follow this training scheme and are devoted to transplanting it to recommendation. \par

With these research effort, the field of self-supervised recommendation recently has demonstrated some promising results showing that mining supervisory signals from stochastic augmentations is desirable \cite{zhou2020s, wu2020self}. However, in contrast to other graph-based tasks, recommendation is distinct because there is widely observed \textsl{homophily} across users and items \cite{Mcpherson2001Birds}. Most existing SSL-based methods conduct the self-discrimination based contrastive learning over the augmented views to learn generalizable representations against the variance in the raw data. Under this scheme, a bijective mapping is built between nodes in two different views, and a given node can just exploit information from itself in another view. Meanwhile, the other nodes are regarded as the negatives that are pushed apart from the given node in the latent space. Obviously, a number of nodes are false negatives which are similar to the given node due to the homophily, and can actually benefit representation learning in the scenario of recommendation if they are recognized as the positives. Conversely, roughly classifying them into the negatives could lead to a performance drop.\par

To tackle this issue, a socially-aware SSL framework which combines the tri-training \cite{zhou2005tri} (multi-view co-training) with SSL is proposed in this paper. For supplementary views that can capture the homophily among users, we resort to social relations which can be another data source that implicitly reflects users' preferences \cite{yu2018adaptive,yu2019generating,chen2020social,yu2020enhance,yin2015dynamic}. Owing to the prevalence of social platforms in the past decade, social relations are now readily accessible in many recommender systems. We exploit the triadic structures in the user-user and user-item interactions to augment two supplementary data views, and socially explain them as profiling users' interests in expanding social circles and sharing desired items to friends, respectively. Given the use-item view which contains users' historical purchases, we have three views that characterize users' preferences from different perspectives and also provide us with a scenario to fuse tri-training and SSL. \par

Tri-training \cite{zhou2005tri} is a popular semi-supervised learning algorithm which exploits unlabeled data using three classifiers. In this work, we employ it to mine self-supervision signals from other users in recommender systems with the multi-view encoding. Technically, we first build three asymmetric graph encoders over the three views, of which two are only for learning user representations and giving pseudo-labels, and another one working on the user-item view also undertakes the task of generating recommendations. Then we dynamically perturb the social network and user-item interaction graph to create an unlabeled example set. Following the regime of tri-training, during each epoch, the encoders over the other two views predict the most probable semantically positive examples in the unlabeled example set for each user in the current view. Then the framework refines the user representations by maximizing the agreement between representations of labeled users in the current view and the example set through the proposed \textit{neighbor-discrimination} based contrastive learning. As all the encoders iteratively improve in this process, the generated pseudo-labels also become more informative, which in turn recursively benefit the encoders again. The recommendation encoder over the user-item view thus becomes stronger in contrast to those only enhanced by the self-discrimination SSL scheme. Since the tri-training operates on the complementary views of the same data sources to learn self-supervision signals, we name it self-supervised tri-training. \par

The major contributions of this paper are summarized as follows:
\begin{itemize}[leftmargin=*]
	\item We propose a general socially-aware self-supervised tri-training framework for recommendation. By unifying the recommendation task and the SSL task under this framework, the recommendation performance can achieve significant gains.
	\item We propose to exploit positive self-supervision signals from other users and develop a neighbor-discrimination based contrastive learning method.
	\item We conduct extensive experiments on multiple real-world datasets to demonstrate the advantages of the proposed SSL framework and investigate the effectiveness of each module in the framework through a comprehensive ablation study. 
\end{itemize} 
The rest of this paper is structured as follows. Section 2 summarizes the related work of recommendation and SSL. Section 3 introduces the proposed framework. The experimental results are reported in Section 4. Finally, Section 5 concludes this paper.

\section{Related Work}
\subsection{Graph Neural Recommendation Models}
Recently, \textit{graph neural networks} (GNNs) \cite{wu2020comprehensive,hamilton2017inductive} have gained considerable attention in the field of recommender systems for their effectiveness in solving graph-related recommendation tasks. Particularly, GCN  \cite{kipf2016semi}, as the prevalent formulation of GNNs which is a first-order approximation of spectral graph convolutions, has driven a multitude of graph neural recommendation models like GCMC \cite{berg2017graph}, NGCF \cite{wang2019neural}, and LightGCN \cite{he2020lightgcn}. The basic idea of these GCN-based models is to exploit the high-order neighbors in the user-item graph by aggregating the embeddings of neighbors to refine the target node's embeddings \cite{wu2020graph}. In addition to these general models, GNNs also empower other recommendation methods working on specific graphs such as SR-GNN \cite{wu2019session} and DHCN \cite{xia2020self} over the session-based graph, and DiffNet \cite{wu2019neural} and MHCN \cite{yu2021self} over the social network. It is worth mentioning that GNNs are often used for social computing as the information spreading in social networks can be well captured by the message passing in GNNs \cite{wu2019neural}. That is the reason why we resort to social networks for self-supervisory signals generated by graph neural encoders.

\begin{figure*}[t]
	\centering
	\includegraphics[width=0.95\textwidth]{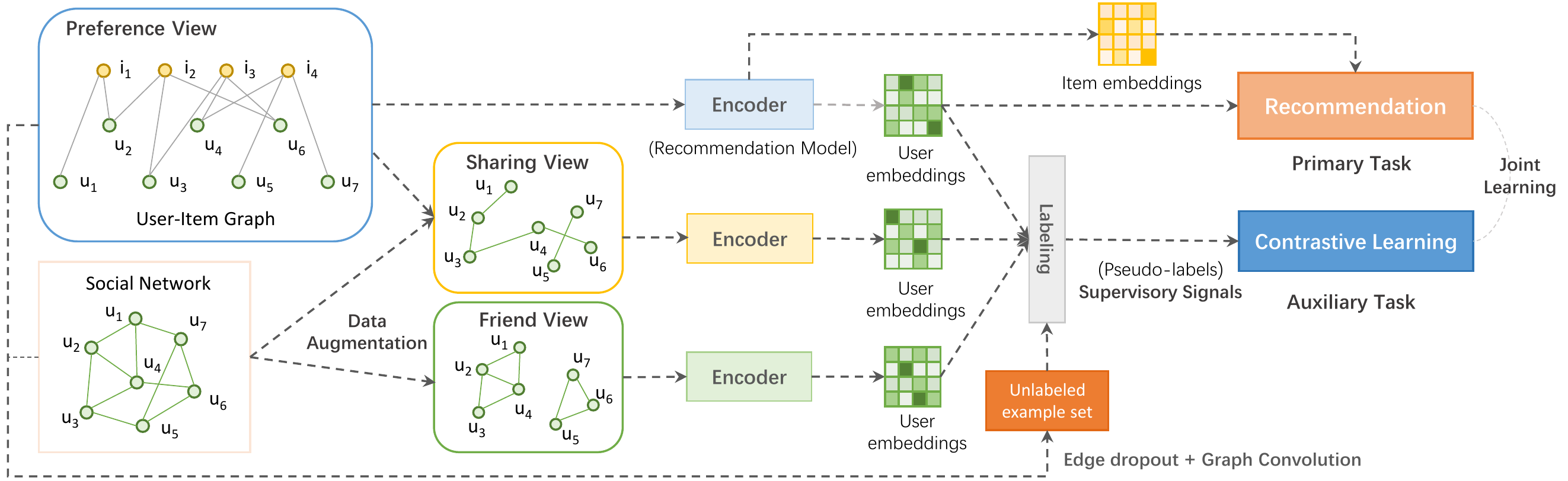}
	\caption{Overview of the proposed self-supervised tri-training framework. }
	\label{figure.1}
\end{figure*}

\subsection{Self-Supervised Learning in RS}
Self-supervised learning \cite{liu2020self} (SSL) is an emerging paradigm to learn with the automatically generated ground-truth samples from the raw data. It was firstly used in visual representation learning and language modeling \cite{hjelm2018learning,bachman2019learning,zhai2019s4l,chen2020simple,he2020momentum} for model pretraining. The recent progress in SSL seeks to harness this flexible learning paradigm for graph representation learning \cite{velickovic2019deep,qiu2020gcc,peng2020graph,sun2019multi}. SSL models over graphs mainly mine self-supervision signals by exploiting the graph structure. The dominant regime of this line of research is graph contrastive learning which contrasts multiple views of the same graph where the incongruent views are built by conducting stochastic augmentations on the raw graph \cite{velickovic2019deep,you2020graph,hassani2020contrastive,qiu2020gcc}. The common types of stochastic augmentations include but are not limited to uniform node/edge dropout, random feature/attribute shuffling, and subgraph sampling using random walk. 
\par
Inspired by the success of graph contrastive learning, there have been some recent works \cite{zhou2020s,yao2020self,wu2020self,ma2020disentangled} which transplant the same idea to the scenario of recommendation. Zhou \textit{et al.} \cite{zhou2020s} devise auxiliary self-supervised objectives by randomly masking attributes of items and skipping items and subsequences of a given sequence for pretraining sequential recommendation model. Yao \textit{et al.} \cite{yao2020self} propose a two-tower DNN architecture with uniform feature masking and dropout for self-supervised item recommendation. Ma \textit{et al.} \cite{ma2020disentangled} mine extra signals for supervision by looking at the longer-term future and reconstruct the future sequence for self-supervision, which adopts feature masking in essence. Wu \textit{et al.} \cite{wu2020self} summarize all the stochastic augmentations on graphs and unify them into a general self-supervised graph learning framework for recommendation. Besides, there are also some studies \cite{xin2020self,yu2021self,sankar2020groupim} refining user representations with mutual information maximization among a set of certain members (e.g. ad hoc groups) for self-supervised recommendation. However, these methods are used for specific situations and cannot be easily generalized to other scenarios.

\section{Proposed Framework}
In this section, we present our \textbf{SE}lf-su\textbf{P}ervised \textbf{T}ri-training framework, called \textbf{SEPT}. The overview of SEPT is illustrated in Fig. \ref{figure.1}. 

\subsection{Preliminaries}
\subsubsection{Notations}
In this paper, we use two graphs as the data sources including the user-item interaction graph $\mathcal{G}_{r}$ and the user social network $\mathcal{G}_{s}$. $\mathcal{U} =\{u_1,u_2, ...,u_m \}$ ($|\mathcal{U}|=m$) denotes the user nodes  across both $\mathcal{G}_{r}$ and $\mathcal{G}_{s}$, and $\mathcal{I} = \{i_1,i_2, ...,i_n\}$ ($|\mathcal{I}|=n$) denotes the item nodes in $\mathcal{G}_{r}$. As we focus on item recommendation, $\bm{R}\in \mathbb{R}^{m\times n}$ is the binary matrix with entries only 0 and 1 that represent user-item interactions in $\mathcal{G}_{r}$. For each entry $(u,i)$ in $\bm{R}$, if user $u$ has consumed/clicked item $i$, $r_{ui}=1$, otherwise $r_{ui}=0$. 
 As for the social relations, we use $\bm{S}\in \mathbb{R}^{m\times m}$ to denote the social adjacency matrix which is binary and symmetric because we work on undirected social networks with bidirectional relations. We use $\bm{P} \in \mathbb{R}^{m\times d}$ and $\bm{Q} \in \mathbb{R}^{n\times d}$ to denote the learned final user and item embeddings for recommendation, respectively. To facilitate the reading, in this paper, matrices appear in bold capital letters and  vectors appear in bold lower letters.

\subsubsection{Tri-Training} Tri-training \cite{zhou2005tri} is a popular semi-supervised learning algorithm which develops from the \textit{co-training} paradigm \cite{blum1998combining} and tackles the problem of determining how to label the unlabeled examples to improve the classifiers. In contrast to the standard co-training algorithm which ideally requires two sufficient, redundant and conditionally independent views of the data samples to build two different classifiers, tri-training is easily applied by lifting the restrictions on training sets. It does not assume sufficient redundancy among the data attributes, and initializes three diverse classifiers upon three different data views generated via bootstrap sampling \cite{efron1994introduction}. Then, in the labeling process of tri-training, for any classifier, an unlabeled example can be labeled for it as long as the other two classifiers agree on the labeling of this example. The generated pseudo-label is then used as the ground-truth to train the corresponding classifier in the next round of labeling. 
\subsection{Data Augmentation}
\subsubsection{View Augmentation}
As has been discussed, there is widely observed homophily in recommender systems. Namely, users and items have many similar counterparts. To capture the homophily for self-supervision, we exploit the user social relations for data augmentation as the social network is often known as a reflection of homophily \cite{Mcpherson2001Birds,Yin2016Adapting} (i.e., users who have similar preferences are more likely to become connected in the social network and vice versa). Since many service providers such as Yelp\footnote{http://www.yelp.com} encourage users to interact with others on their platforms, it provides their recommender systems with opportunities to leverage abundant social relations. However, as social relations are inherently noisy \cite{yu2020enhance,yu2018adaptive}, for accurate supplementary supervisory information, SEPT only utilizes the reliable social relations by exploiting the ubiquitous triadic closure \cite{huang2014mining} among users. In a socially-ware recommender system, by aligning the user-item interaction graph $\mathcal{G}_{r}$ and the social network $\mathcal{G}_{s}$, we can readily get two types of triangles: three users socially connected with each other (e.g. $u_{1}, u_{2}$ and $u_{4}$ in Fig. \ref{figure.1}) and two socially connected users with the same purchased item (e.g. $u_{1}, u_{2}$ and $i_{1}$ in Fig. \ref{figure.1}). The former is socially explained as profiling users' interests in expanding social circles, and the latter is characterizing users' interests in sharing desired items with their friends. It is straightforward to regard the triangles as strengthened ties because if two persons in real life have mutual friends or common interests, they are more likely to have a close relationship. 
\par
Following our previous work \cite{yu2021self}, the mentioned two types of triangles can be efficiently extracted in the form of matrix multiplication. Let $\bm{A}_{f}\in\mathbb{R}^{m\times m}$ and $\bm{A}_{s}\in\mathbb{R}^{m\times m}$ denote the adjacency matrices of the users involved in these two types of triangular relations. They can be calculated by:
\begin{equation}
\bm{A}_{f} = (\bm{S}\bm{S})\odot \bm{S},\ \ \bm{A}_{s} = (\bm{R}\bm{R}^{\top})\odot \bm{S}.
\end{equation}
The multiplication $\bm{S}\bm{S}$ ($\bm{R}\bm{R}^{\top}$) accumulates the paths connecting two user via shared friends (items), and the Hadamard product $\odot\bm{S}$ makes these paths into triangles. Since both $\bm{S}$ and $\bm{R}$ are sparse matrices, the calculation is not time-consuming.  The operation $\odot \bm{S}$ ensures that the relations in $\bm{A}_{f}$ and $\bm{A}_{s}$ are subsets of the relations in $\bm{S}$. As $\bm{A}_{f}$ and $\bm{A}_{s}$ are not binary matrices, Eq. (1) can be seen a special case of bootstrap sampling on $\bm{S}$ with the complementary information from $\bm{R}$. Given $\bm{A}_{f}$ and $\bm{A}_{s}$ as the augmentation of $\bm{S}$ and $\bm{R}$, we have three views that characterize users' preferences from different perspectives and also provide us with a scenario to fuse tri-training and SSL. To facilitate the understanding, we name the view over the user-item interaction graph \textit{preference view}, the view over the triangular social relations \textit{friend view}, and another one \textit{sharing view}, which are represented by $\bm{R}$, $\bm{A}_{f}$, and $\bm{A}_{s}$, respectively. 
\subsubsection{Unlabeled Example Set}
To conduct tri-training, an unlabeled example set is required. We follow existing works \cite{you2020graph,wu2020self} to perturb the raw graph with edge dropout at a certain probability $\rho$ to create a corrupted graph from where the learned user presentations are used as the unlabeled examples. This process can be formulated as:
\begin{equation}
\mathcal{\tilde{\mathcal{G}}} = (\mathcal{N}_{r}\cup \mathcal{N}_{s}, \bm{m}\odot(\mathcal{E}_{r}\cup \mathcal{E}_{s})),
\end{equation}
where $\mathcal{N}_{r}$ and $\mathcal{N}_{s}$ are nodes, $\mathcal{E}_{r}$ and $\mathcal{E}_{s}$ are edges in $\mathcal{G}_{r}$ and $\mathcal{G}_{s}$, and $\bm{m}\in\{0,1\}^{|\mathcal{E}_{r}\cup \mathcal{E}_{s}|}$ is the mask vector to drop edges. Herein we perturb both $\mathcal{G}_{r}$ and $\mathcal{G}_{s}$ instead of $\mathcal{G}_{r}$ only, because the social information is included in the aforementioned two augmented views. For integrated self-supervision signals, perturbing the joint graph is necessary. 

\subsection{SEPT: Self-Supervised Tri-Training}
\subsubsection{Architecture}
With the augmented views and the unlabeled example set, we follow the setting of tri-training to build three encoders. Architecturally, the proposed self-supervised training framework can be model-agnostic so as to boost a multitude of graph neural recommendation models. But for a concrete framework which can be easily followed, we adopt LightGCN \cite{he2020lightgcn} as the basic structure of the encoders due to its simplicity. The general form of encoders is defined as follows:
\begin{equation}
\bm{Z}=H(\bm{E},\mathcal{V}),
\end{equation}
where $H$ is the encoder, $\bm{Z}\in\mathbb{R}^{m\times d}$ or $\mathbb{R}^{(m+n)\times d}$ denotes the final representation of nodes, $\bm{E}$ of the same size denotes the initial node embeddings which are the bottom shared by the three encoders, and $\mathcal{V} \in \{\bm{R},\bm{A}_{s},\bm{A}_{f}\}$ is any of the three views. It should be noted that, unlike the vanilla tri-training, SEPT is asymmetric. The two encoders $H_{f}$ and $H_{s}$ that work on the \textit{friend view} and \textit{sharing view} are only in charge of learning user representations through graph convolution and giving pseudo-labels, while the encoder $H_{r}$ working on the \textit{preference view} also undertakes the task of generating recommendations and thus learns both user and item representations (shown in Fig. \ref{figure.1}). Let $H_{r}$ be the dominant encoder (recommendation model), and $H_{f}$ and $H_{s}$ be the auxiliary encoders. Theoretically, given a concrete $H_{r}$ like LightGCN \cite{he2020lightgcn}, there should be the optimal structures of $H_{f}$ and $H_{s}$. However, exploring the optimal structures of the auxiliary encoders is out of the scope of this paper. For simplicity, we assign the same structure to $H_{f}$ and $H_{s}$. Besides, to learn representations of the unlabeled examples from the perturbed graph $\tilde{\mathcal{G}}$, another encoder is required, but it is only for graph convolution. All the encoders share the bottom embeddings $\bm{E}$ and are built over different views with the LightGCN structure.  
\subsubsection{Constructing Self-Supervision Signals}
By performing graph convolution over the three views, the encoders learn three groups of user representations. As each view reflects a different aspect of the user preference, it is natural to seek supervisory information from the other two views to improve the encoder of the current view. Given a user, we predict its semantically positive examples in the unlabeled example set using the user representations from the other two views. Taking user $u$ in the preference view as an instance, the labeling is formulated as:
\begin{equation}
\begin{split}
\bm{y}_{u+}^{s} = Softmax(\phi(\tilde{\bm{Z}},\bm{z}_{u}^{s})), \ \ \bm{y}_{u+}^{f} = Softmax(\phi(\tilde{\bm{Z}},\bm{z}_{u}^{f})), 
\end{split}
\end{equation}
where $\phi$ is the \textit{cosine} operation, $\bm{z}_{u}^{s}$ and $\bm{z}_{u}^{f}$ are the representations of user $u$ learned by $H_{s}$ and $H_{f}$, respectively, $\tilde{\bm{Z}}$ is the representations of users in the unlabeled example set obtained through graph convolution, and $\bm{y}_{u+}^{s}$ and $\bm{y}_{u+}^{f}$ denote the predicted probability of each user being the semantically positive example of user $u$ in the corresponding views. \par
Under the scheme of tri-training, to avoid noisy examples, only if both $H_{s}$ and $H_{f}$ agree on the labeling of a user being the positive sample, and then the user can be labeled for $H_{r}$. We obey this rule and add up the predicted probabilities from the two views and obtain:
\begin{equation}
\bm{y}_{u+}^{r} = \frac{1}{2}(\bm{y}_{u+}^{s}+\bm{y}_{u+}^{f}).
\end{equation}
With the probabilities, we can select $K$ positive samples with the highest confidence. This process can be formulated as:
\begin{equation}
\mathcal{P}_{u+}^{r} = \{\tilde{\bm{Z}}_{k}\ |\ k\in \text{Top-}K(\bm{y}_{u+}^{r}),\ \tilde{\bm{Z}}\sim\mathcal{\tilde{\mathcal{G}}} \}.
\end{equation}
In each iteration, $\mathcal{\tilde{\mathcal{G}}}$ is reconstructed with the random edge dropout for varying user representations. SEPT dynamically generates positive pseudo-labels over this data augmentation for each user in every view. Then these labels are used as the supervisory signals to refine the shared bottom representations. 
\subsubsection{Contrastive Learning}
Having the generated pseudo-labels, we develop the \textit{neighbor-discrimination} contrastive learning method to fulfill self-supervision in SEPT.\par
Given a certain user, we encourage the consistency between his node representation and the labeled user representations from $\mathcal{P}_{u+}$, and minimize the agreement between his node representation and the unlabeled user representations. The idea of the neighbor-discrimination is that, given a certain user in the current view, the positive pseudo-labels semantically represent his neighbors or potential neighbors in the other two views, then we should also bring these positive pairs together in the current view due to the homophily across different views. And this can be achieved through the \textit{neighbor-discrimination} contrastive learning. Formally, we follow the previous studies \cite{wu2020self,chen2020simple} to adopt InfoNCE \cite{hjelm2018learning}, which is effective in mutual information estimation, as our learning objective to maximize the agreement between positive pairs and minimize that of negative pairs:
\begin{equation}
\mathcal{L}_{ssl}=-\mathbb{E}\sum_{v\in\{r,s,f\}}\left[\log \frac{\sum_{p \in \mathcal{P}_{u+}^{v}} \psi(z_{u}^{v},\tilde{z}_{p})}{\sum_{p \in \mathcal{P}_{u+}^{v}}  \psi(z_{u}^{v},\tilde{z}_{p})+\sum_{j \in U/\mathcal{P}_{u+}^{v}} \psi(z_{u}^{v},\tilde{z}_{j})}\right]
\end{equation}
where $\psi(z_{u}^{v},\tilde{z}_{p})=\exp\left(\phi(z_{u}^{v} \cdot \tilde{z}_{p}) / \tau\right)$, $\phi(\cdot): \mathbb{R}^{d} \times \mathbb{R}^{d} \longmapsto \mathbb{R}$ is the discriminator function that takes two vectors as the input and then scores the agreement between them, and $\tau$ is the temperature to amplify the effect of discrimination ($\tau=0.1$ is the best in our implementation). We simply implement the discriminator by applying the cosine operation. Compared with the self-discrimination, the neighbor-discrimination leverages the supervisory signals from the other users. When only one positive example is used and if the user itself in $\tilde{Z}$ has the highest confidence in $\bm{y}_{u+}$, the neighbor-discrimination degenerates to the self-discrimination. So, the self-discrimination can be seen as a special case of the neighbor-discrimination. But when a sufficient number of positive examples are used, these two methods could also be simultaneously adopted because the user itself in $\tilde{Z}$ is often highly likely to be in the Top-\textit{K} similar examples $\mathcal{P}_{u+}$. With the training proceeding, the encoders iteratively improve to generate evolving pseudo-labels, which in turn recursively benefit the encoders again.\par
Compared with the vanilla tri-training, it is worth noting that in SEPT, we do not add the pseudo-labels into the adjacency matrices for subsequent graph convolution during training. Instead, we adopt a soft and flexible way to guide the user representations via mutual information maximization, which is distinct from the vanilla tri-training that adds the pseudo-labels to the training set for next-round training. The benefits of this modeling are two-fold. Firstly, adding pseudo-labels leads to reconstruction of the adjacency matrices after each iteration, which is time-consuming; secondly, the pseudo-labels generated at the early stage might not be informative; repeatedly using them would mislead the framework.
\subsubsection{Optimization} 
The learning of SEPT consists of two tasks: recommendation and the neighbor-discrimination based contrastive learning. Let $\mathcal{L}_{r}$ be the BPR pairwise loss function \cite{rendle2009bpr} which is defined as:
\begin{equation}
\mathcal{L}_{r}=\sum_{i \in \mathcal{I}(u), j \notin \mathcal{I}(u)}-\log \sigma(\hat{r}_{ui}-\hat{r}_{uj})+\lambda\|\bm{E}\|_{2}^{2},
\end{equation}
where $\mathcal{I}(u)$ is the item set that user $u$ has interacted with, $\hat{r}_{ui}=\bm{P}^{\top}_{u}\bm{Q}_{i}$, $\bm{P}$ and $\bm{Q}$ are obtained by splitting $\bm{Z}^{r}$, and $\lambda$ is the coefficient controlling the $L_{2}$ regularization.
The training of SEPT proceeds in two stages: \textit{initialization} and \textit{joint learning}.  To start with, we warm up the framework with the recommendation task by optimizing $\mathcal{L}_{r}$. Once trained with $\mathcal{L}_{r}$, the shared bottom $\bm{E}$ has gained far stronger representations than randomly initialized embeddings. The self-supervised tri-training then proceeds as described in Eq. (4) - (7), acting as an auxiliary task which is unified into a joint learning objective to enhance the performance of the recommendation task. The overall objective of the joint learning is defined as:
\begin{equation}
\mathcal{L}=\mathcal{L}_{r}+\beta\mathcal{L}_{ssl},
\end{equation}
where $\beta$ is a hyper-parameter used to control the magnitude of the self-supervised tri-training. The overall process of SEPT is presented in Algorithm \ref{alg:Framework}.

\begin{algorithm}[t]
	\caption{The running process of SEPT}
	\LinesNumbered 
	\label{alg:Framework}
	
	\KwIn{Bidirectional social relations $\mathbf{S}$, User feedback $\mathbf{R}$}, and randomly initialized node embeddings $\bm{E}$ \;
	\KwOut{Recommendation lists} 	   
	Pretraining with $\mathcal{L}_{r}$ in Eq. (8)\;
	View augmentation with Eq. (1)\;
	\For {each iteration}{
		Construct $\tilde{\mathcal{G}}$ and obtain the unlabeled example set through graph convolution\;
		\For {each batch}{
			Randomly select $c$ users from $\bm{\tilde{Z}}$ to be labeled\;
			\For {each user $u$}{
				Predict the probabilities of the $c$ users being the semantically positive examples in different views with Eq. (4) - (5)\;
				Obtain Top-\textit{K} positive examples with Eq. (6)\;				
			}
		Jointly optimize the overall objective in Eq. (9)\;
		}
   }
\end{algorithm}

\subsection{Discussions}
\subsubsection{Connection with Social Regularization} Social recommendation \cite{yu2021self,yu2020enhance,yin2015dynamic} integrates social relations into recommender systems to address the data sparsity issue. A common idea of social recommendation is to regularize user representations by minimizing the euclidean distance between socially connected users, which is termed \textit{social regularization} \cite{ma2011recommender}. Although the proposed SEPT also leverages socially-aware supervisory signals to refine user representations, it is distinct from the social regularization. The differences are also two-fold. Firstly, the social regularization is a static process which is always performed on the socially connected users, whereas the neighbor-discrimination is dynamic and iteratively improves the supervisory signals imposed on uncertain users; secondly, negative social relations (dislike) cannot be readily retrieved in social recommendation, and hence the social regularization can only keep socially connected users close. But SEPT can also pushes users who are not semantically positive in the three views apart.
\subsubsection{Complexity} Architecturally, SEPT can be model-agnostic, and its complexity mainly depends on the structure of the used encoders. In this paper, we present a LightGCN-based architecture. Given $\mathcal{O}(|\bm{R}|d)$ as the time complexity of the recommendation encoder for graph convolution, the total complexity for the graph convolution is less than $4\mathcal{O}(|\bm{R}|d)$ because $\bm{A}_{f}$,  $\bm{A}_{s}$, and $\tilde{\mathcal{G}}$ are usually sparser than $\bm{R}$. Another cost  comes from the Top-\textit{K} operation of the labeling process in Eq. (6), which usually requires $\mathcal{O}(m\log(K))$ by using the max heap. To reduce the cost and speed up training, in each batch for training, only $c$ ($c\ll m$, e.g. 1000) users in a batch are randomly selected and being the unlabeled example set of the pseudo-labels, and this sampling method can also prevent overfitting. The complexity of the neighbor-discrimination based contrastive learning is $\mathcal{O}(cd)$.  

	\begin{table}[ht]
	\small
	\renewcommand\arraystretch{1.1}
	\caption{Dataset Statistics}
	
	\label{Table:1}
	\begin{center}
		\begin{tabular}{c|ccccc}
			\hline
			Dataset&\#User & \#Item &  \#Feedback &  \#Relation & Density\\ \hline
			\hline
			Last.fm &1,892 &  17,632 & 92,834 & 25,434  & 0.28\%\\
			Douban-Book & 13,024 & 22,347 & 792,062 &169,150 & 0.27\%\\
			Yelp&19,539  &21,266 & 450,884 &864,157 & 0.11\%\\
			\hline
		\end{tabular}
	\end{center}
\end{table}

\begin{table*}[t]
	\small
	\caption{Performance improvements brought by SEPT on the three datasets.}
	\label{Table:2}
	\renewcommand\arraystretch{1.1}
	\begin{center}
		{
			\begin{tabular}{*{11}{c}}
				\toprule
				\multicolumn{2}{c}{\multirow{2}{*}{Method}}&
				\multicolumn{3}{c}{Last.fm} & \multicolumn{3}{c}{Douban-Book} & \multicolumn{3}{c}{Yelp} \cr
				\cmidrule(lr){3-5}\cmidrule(lr){6-8}\cmidrule(lr){9-11} && Prec@10 & Rec@10 & NDCG@10 &  Prec@10 & Rec@10 & NDCG@10 &  Prec@10 & Rec@10 & NDCG@10  \\ \hline				
				
				\multirow{2}{*}{\textbf{1-Layer}} &LightGCN &17.517 & 17.463 & 21.226 & 6.712 & 8.248 & 9.584 & 2.433 & 6.083 & 4.688  \\
				
				&\multirow{2}{*}{\textbf{SEPT}}     &18.266 & 18.503 & 21.985 & 6.931 &   8.984   &   10.116   &   2.620   &  6.738    & 5.233  \\
				
				&&\footnotesize{$\uparrow$ 4.275\%} & \footnotesize{$\uparrow$ 5.955\%} & \footnotesize{$\uparrow$ 3.575\%} & \footnotesize{$\uparrow$ 3.262\%} & \footnotesize{$\uparrow$ 8.923\%} & \footnotesize{$\uparrow$ 5.551\%}   &\footnotesize{$\uparrow$ 7.685\%} & \footnotesize{$\uparrow$ 10.767\%} & \footnotesize{$\uparrow$ 11.625\%}   \\
				\hline		
				
				\multirow{2}{*}{\textbf{2-Layer}} & LightGCN & 19.205 & 19.480 & 23.392 & 7.650 &   10.024   &   11.348 & 2.641 & 6.791 & 5.235  \\
				
				&\multirow{2}{*}{\textbf{SEPT}}     &20.191 & 20.488 & 24.507 & 8.383 &   10.810   &   12.635   &   2.901    & 7.460& 5.776  \\
				
				&&\footnotesize{$\uparrow$ 5.134\%} & \footnotesize{$\uparrow$ 5.174\%} & \footnotesize{$\uparrow$ 4.766\%} & \footnotesize{$\uparrow$ 9.582\%} & \footnotesize{$\uparrow$ 7.841\%} & \footnotesize{$\uparrow$ 11.341\%}   &\footnotesize{$\uparrow$ 9.844\%} & \footnotesize{$\uparrow$ 9.851\%} & \footnotesize{$\uparrow$ 10.334\%}   \\
				
				\hline				
				\multirow{2}{*}{\textbf{3-Layer}} & LightGCN  & 18.827 & 19.234 & 22.904 & 7.275 & 9.599 & 10.836& 2.426 & 6.130 & 4.703  \\
				
				&\multirow{2}{*}{\textbf{SEPT}}     &19.082 & 19.363 & 23.175 &   7.868   &   10.458   &   11.928   &  2.635    & 6.807& 5.250  \\
				
				&&\footnotesize{$\uparrow$ 1.354\%} & \footnotesize{$\uparrow$ 0.671\%} & \footnotesize{$\uparrow$ 1.183\%} & \footnotesize{$\uparrow$ 8.151\%} & \footnotesize{$\uparrow$ 8.948\%} & \footnotesize{$\uparrow$ 10.077\%}   &\footnotesize{$\uparrow$ 8.615\%} & \footnotesize{$\uparrow$ 11.044\%} & \footnotesize{$\uparrow$ 11.631\%}   \\		
				
				\bottomrule
		\end{tabular}}
	\end{center}
\end{table*}

\section{Experimental Results}
\subsection{Experimental Settings} 
\noindent\textbf{Datasets.} Three real-world datasets: Last.fm\footnote{http://files.grouplens.org/datasets/hetrec2011/}, Douban-Book\footnote{https://github.com/librahu/HIN-Datasets-for-Recommendation-and-Network-Embedding}, and Yelp \footnote{https://github.com/Coder-Yu/QRec} are used in our experiments to evaluate SEPT. As SEPT aims to improve Top-N recommendation, we follow the convention in previous research \cite{yu2021self,yu2020enhance} to leave out ratings less than 4 in the dataset of Douban-Book which consists of explicit ratings with a 1-5 rating scale, and assign 1 to the rest. The statistics of the datasets is shown in Table 1. For precise assessment, 5-fold cross-validation is conducted in all the experiments and the average results are presented. \par
\noindent\textbf{Baselines.} Three recent graph neural recommendation models are compared with SEPT to test the effectiveness of the self-supervised tri-training for recommendation:
\begin{itemize}[leftmargin=*]
	\item \textbf{LightGCN} \cite{he2020lightgcn} is a GCN-based general recommendation model that leverages the user-item proximity to learn node representations and generate recommendations, which is reported as the state-of-the-art.
	\item \textbf{DiffNet++} \cite{wu2020diffnet++} is a recent GCN-based social recommendation method that models the recursive dynamic social diffusion in both the user and item spaces.
	\item \textbf{MHCN} \cite{yu2021self} is a latest hypergraph convolutional network-based social recommendation method that models the complex correlations among users with hyperedges to improve recommendation performance.
\end{itemize} 
LightGCN \cite{he2020lightgcn} is the basic encoder in SEPT. Investigating the performance of LightGCN and SEPT is essential. Since LightGCN is a widely acknowledged SOTA baseline reported in many recent papers \cite{yu2021self,wu2020self}, we do not compare SEPT with other weak baselines such as NGCF \cite{wang2019neural}, GCMC \cite{berg2017graph}, and BPR \cite{rendle2009bpr}.
Two strong social recommendation models are also compared to SEPT to verify that the self-supervised tri-training, rather than the use of social relations, is the main driving force of the performance improvements.

\noindent\textbf{Metrics.} To evaluate all the methods, we first perform item ranking on all the candidate items. Then two relevancy-based metrics \emph{Precision@10} and \emph{Recall@10} and one ranking-based metric \emph{NDCG@10} are calculated on the truncated recommendation lists, and the values are presented in percentage.\\
\noindent\textbf{Settings.} For a fair comparison, we refer to the best parameter settings reported in the original papers of the baselines and then fine tune all the hyperparameters of the baselines to ensure the best performance of them. As for the general settings of all the methods, we empirically set the dimension of latent factors (embeddings) to 50, the regularization parameter $\lambda$ to 0.001, and the batch size to 2000. In section 4.4, we investigate the parameter sensitivity of SEPT, and the best parameters are used in section 4.2 and 4.3. We use Adam to optimize all these models with an initial learning rate 0.001. 
\par

\begin{table*}[h]
	\caption{Performance comparison with social recommendation models on three datasets.}
	\label{Table:3}
	\renewcommand\arraystretch{1.1}
	\begin{center}
		{
			\begin{tabular}{*{10}{c}}
				\toprule
				\multirow{2}{*}{Method} &
				\multicolumn{3}{c}{Last.fm} & \multicolumn{3}{c}{Douban-Book} & \multicolumn{3}{c}{Yelp} \cr
				\cmidrule(lr){2-4}\cmidrule(lr){5-7}\cmidrule(lr){8-10} & Prec@10 & Rec@10 & NDCG@10 &  Prec@10 & Rec@10 & NDCG@10 &  Prec@10 & Rec@10 & NDCG@10  \\ \hline				
				
				DiffNet++ &18.485 & 18.737 & 22.310 & 7.250 & 9.511 & 9.591 & 2.480 & 6.354 & 4.833  \\

				MHCN  & 19.625 & 19.945 & 23.834 & 7.718 &   10.113   &   11.540 & 2.751 & 6.862 & 5.356  \\
				
				$S^{2}$-MHCN  & 20.052 & 20.375 & 24.395 & 8.083 &   10.402   &   12.136 & 3.003 & 7.885 & 6.061  \\					
				
				\hline								
				\multirow{2}{*}{\textbf{SEPT}}     &20.191 & 20.488 & 24.507  &    8.383 &   10.810   &   12.635   &  2.901    & 7.460& 5.776  \\
				
				&\footnotesize{$\uparrow$ 0.693\%} & \footnotesize{$\uparrow$ 0.554\%} & \footnotesize{$\uparrow$ 0.459\%} & \footnotesize{$\uparrow$ 3.711\%} & \footnotesize{$\uparrow$ 3.922\%} & \footnotesize{$\uparrow$ 4.111\%}   &\footnotesize{$\downarrow$ 3.396\%} & \footnotesize{$\downarrow$ 5.389\%} & \footnotesize{$\downarrow$ 4.702\%}   \\		
				
				\bottomrule
		\end{tabular}}
	\end{center}
\end{table*}

\subsection{Overall Performance Comparison}
In this part, we validate if SEPT can improve recommendation. The performance comparisons are shown in Table \ref{Table:2} and \ref{Table:3}. We conduct experiments with different layer numbers in Table \ref{Table:2}. In Table \ref{Table:3}, a two-layer setting is adopted for all the methods because they all reach their best performance on the used datasets under this setting. The performance improvement (drop) marked by $\uparrow$ ($\downarrow$) is calculated by using the performance difference to divide the subtrahend. According to the results, we can draw the following observations and conclusions: 
	\begin{itemize}[leftmargin=*]
	\item Under all the different layer settings, SEPT can significantly boost LightGCN. Particularly, on the sparser datasets: Douban-Book and Yelp, the improvements get higher. The maximum improvement can even reach 11\%. This can be an evidence that demonstrates the effectiveness of self-supervised learning. Besides, although both LightGCN and SEPT suffer the over-smoothed problem when the layer number is 3, SEPT can still outperform LightGCN. We think the possible reason is that contrastive learning can, to some degree, alleviate the over-smooth problem because the dynamically generated unlabeled examples provide sufficient data variance.
\end{itemize}
In addition to the comparison with LightGCN, we also compare SEPT with social recommendation models to validate if the self-supervised tri-training rather than social relations primarily promote the recommendation performance. Since MHCN is also built upon LightGCN, comparing these two models can be more informative. Besides, $S^{2}$-MHCN, which is the self-supervised variant of MHCN is also compared. The improvements (drops) are calculated by comparing the results of SEPT and $S^{2}$-MHCN. According to the results in Table \ref{Table:3}, we make the following observations and conclusions:
	\begin{itemize}[leftmargin=*]
	\item Although integrating social relations into graph neural models are helpful (comparing MHCN with LightGCN), learning under the scheme of SEPT can achieve more performance gains (comparing SEPT with MHCN). DiffNet++ is uncompetitive compared with the other three methods. Its failure can be attributed to its redundant and useless parameters and operations \cite{he2020lightgcn}. On both LastFM and Douban-Book, SEPT outperforms $S^{2}$-MHCN. On Yelp, $S^{2}$-MHCN exhibits better performance than SEPT does. The superiority of SEPT and $S^{2}$-MHCN demonstrates that self-supervised learning holds vast capability for improving recommendation. In addition, SEPT does not need to learn other parameters except the bottom embeddings, whereas there are a number of other parameters that $S^{2}$-MHCN needs to learn. Meanwhile, SEPT runs much faster than $S^{2}$-MHCN does in our experiments, which makes it more competitive even that it is beaten by $S^{2}$-MHCN on Yelp by a small margin. 
\end{itemize}

\begin{figure}[t]
	\centering
	\includegraphics[width=.5\textwidth]{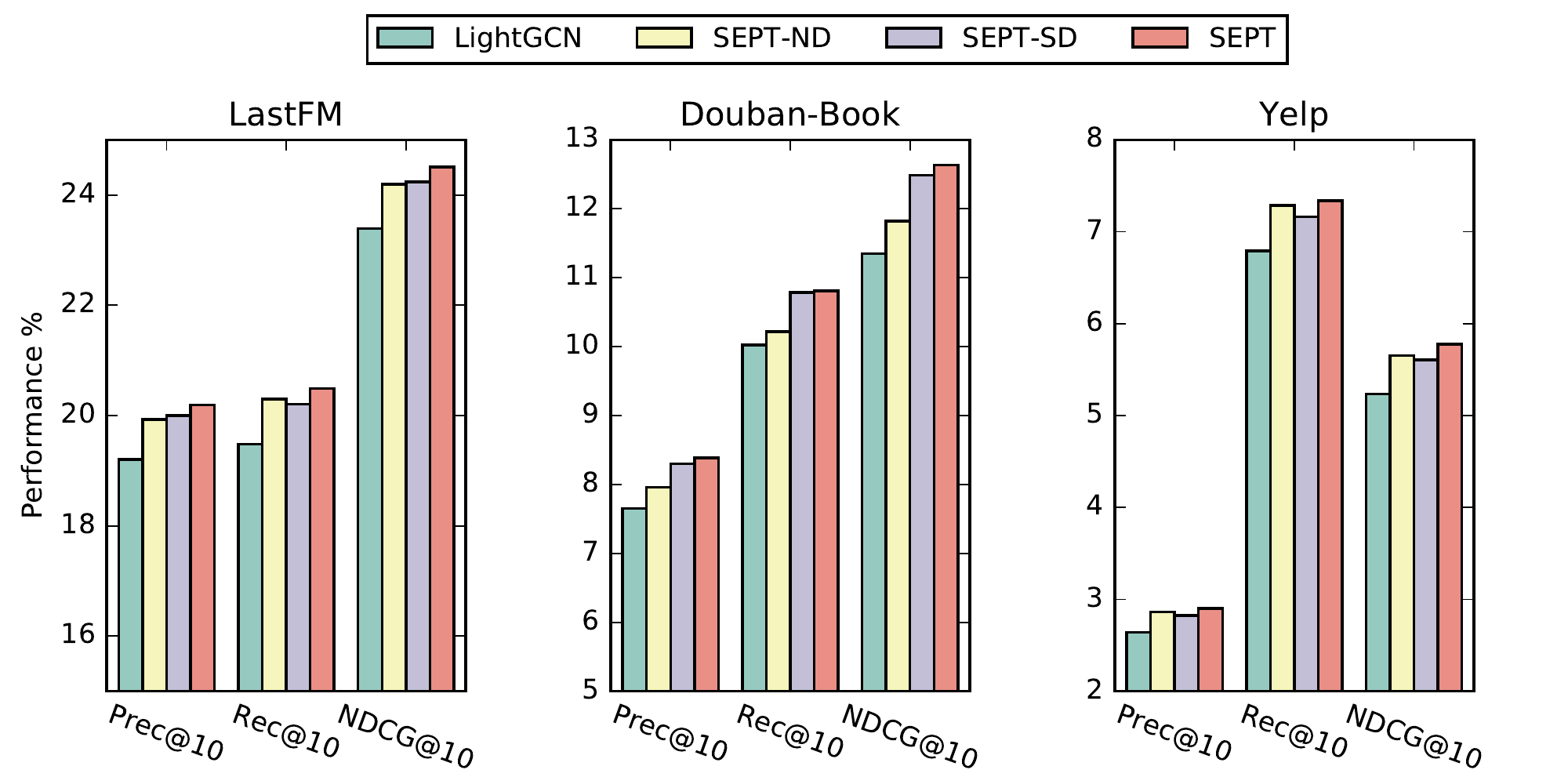}
	\caption{Comparisons between self-discrimination and neighbor-discrimination. }
	\label{figure.2}
	\vspace{-10px}		
\end{figure}

\subsection{Self-Discrimination \textit{v.s.} Neighbor-Discrimination}
In SEPT, the generated positive examples can include both the user itself and other users in the unlabeled example set. It is not clear which part contributes more to the recommendation performance. In this part, we investigate the self-discrimination and the neighbor-discrimination without the user itself being the positive example. For convenience, we use SEPT-SD to denote the self-discrimination, and SEPT-ND to denote the latter. It also should be mentioned that, for SEPT-ND only, a small $\beta=0.001$ can lead to the best performance on all the datasets. A two-layer setting is used in this case.\par
\begin{figure}[t]
	\centering
	\includegraphics[width=.5\textwidth]{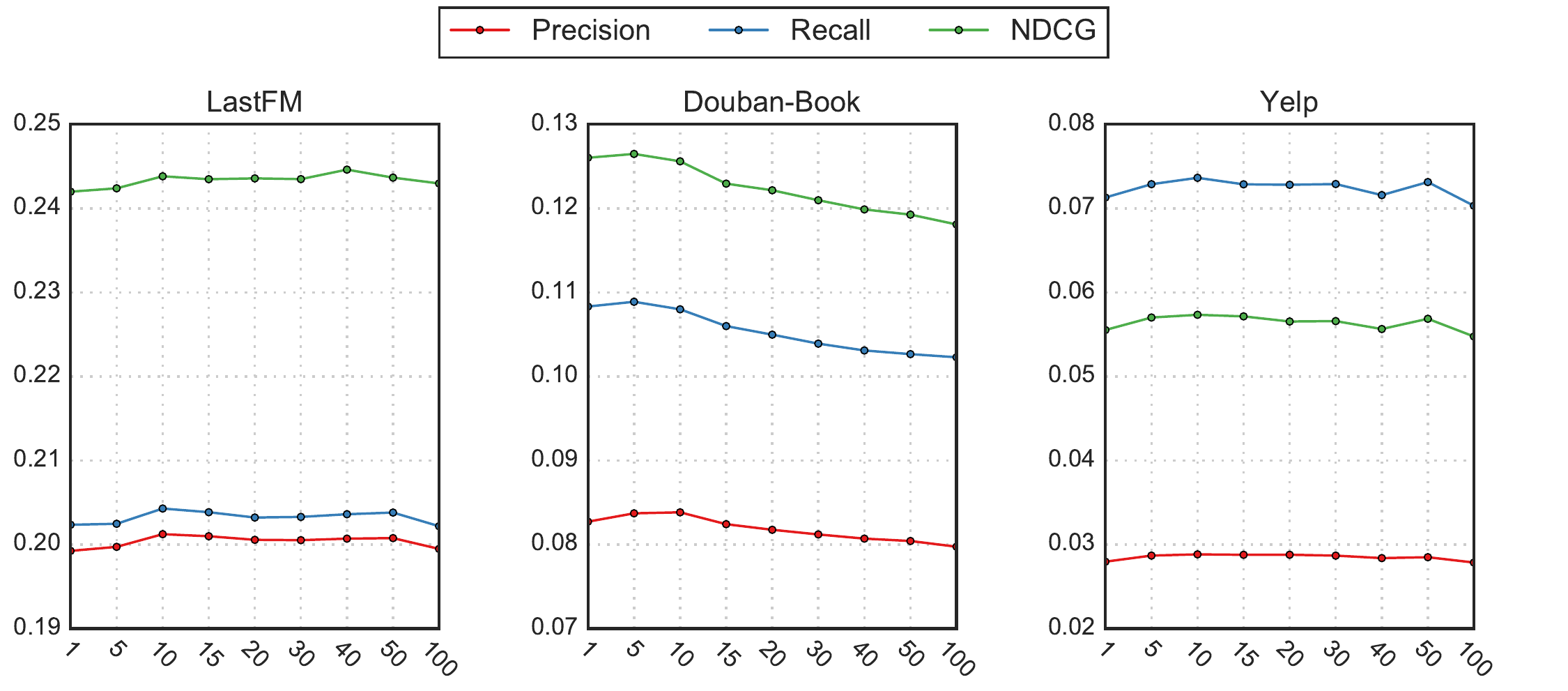}
	\caption{Influence of the number of used positive examples.}
	\label{figure.5}
\end{figure}

According to Fig. \ref{figure.2}, we can observe that both SEPT-SD and SEPT-ND exhibit better performances than LightGCN does, which proves that both the supervisory signals from the user itself and other users can benefit a self-supervised recommendation model. Our claim about the self-supervision signals from other users is validated. Besides, the importance of the self-discrimination and the neighbor-discrimination varies from dataset to dataset. On LastFM, they almost contribute equally. On Douban-Book, self-discrimination shows much more importance. On Yelp, neighbor-discrimination is more effective. This phenomenon can be explained by Fig. \ref{figure.5}. With the increase of the used positive examples, we see that the performance of SEPT almost remains stable on LastFM and Yelp but gradually declines on Douban-Book. We guess that there is widely observed homophily in LastFM and Yelp, so a large number of users share similar preferences, which can be the high-quality positive examples in these two datasets. However, users in Douban-Book may have more diverse interests, which results in the quality drop when the number of used positive examples increases.

\begin{figure}[t]
	\centering
	\includegraphics[width=.5\textwidth]{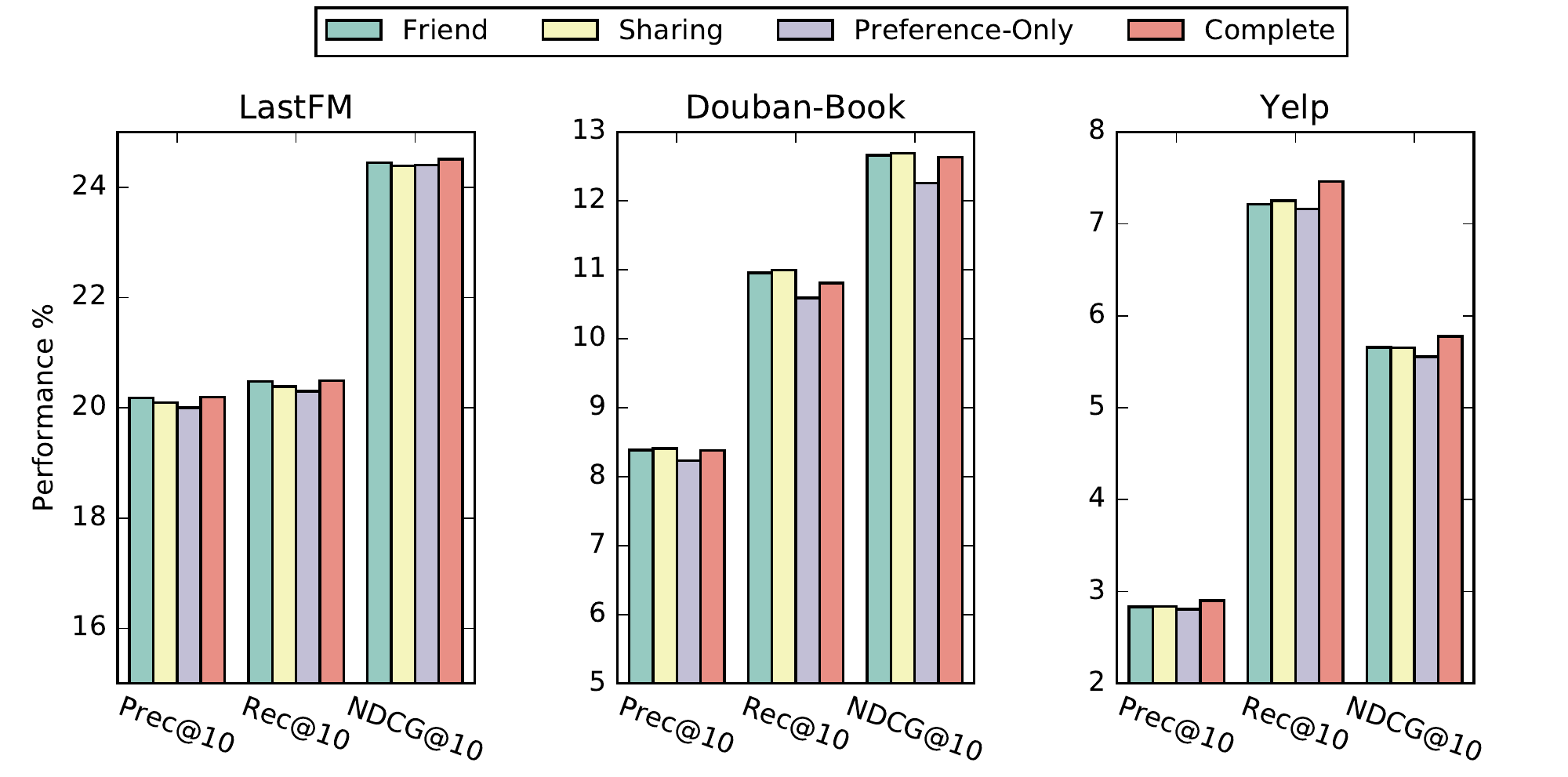}
	\caption{Contributions of each component in SEPT. }
	\label{figure.3}
	\vspace{-10px}		
\end{figure}
\subsection{View Study}
In SEPT, we build two augmented views to conduct tri-training for mining supervisory signals. In this part, we ablate the framework to investigate the contribution of each view. A two-layer setting is used in this case. In Fig. \ref{figure.3}, `Friend' or `Sharing' means that the corresponding view is detached. When only two views are used, SEPT degenerates to the self-supervised co-training. `Preference-Only' means that only the preference view is used. In this case, SEPT further degenerates to the self-training. \par
From Fig. \ref{figure.3}, we can observe that on both LastFM and Yelp, all the views contribute, whereas on Douban-Book, the self-supervised co-training setting achieves the best performance. Moreover, when only the preference view is used, SEPT shows lower performance but it is still better than that of LightGCN. With the decrease of used number of views, the performance of SEPT slightly declines on LastFM, and an obvious performance drop is observed on Yelp. On Douban-Book, the performance firstly gets a slight rise and then declines obviously when there is only one view. The results demonstrate that, under the semi-supervised setting, even a single view can generate desirable self-supervised signals, which is encouraging since social relations or other side information are not always accessible in some situations. Besides,  increasing the used number of views may bring more performance gains, but it is not absolutely right.

\begin{figure}[t]
	\centering
	\includegraphics[width=.5\textwidth]{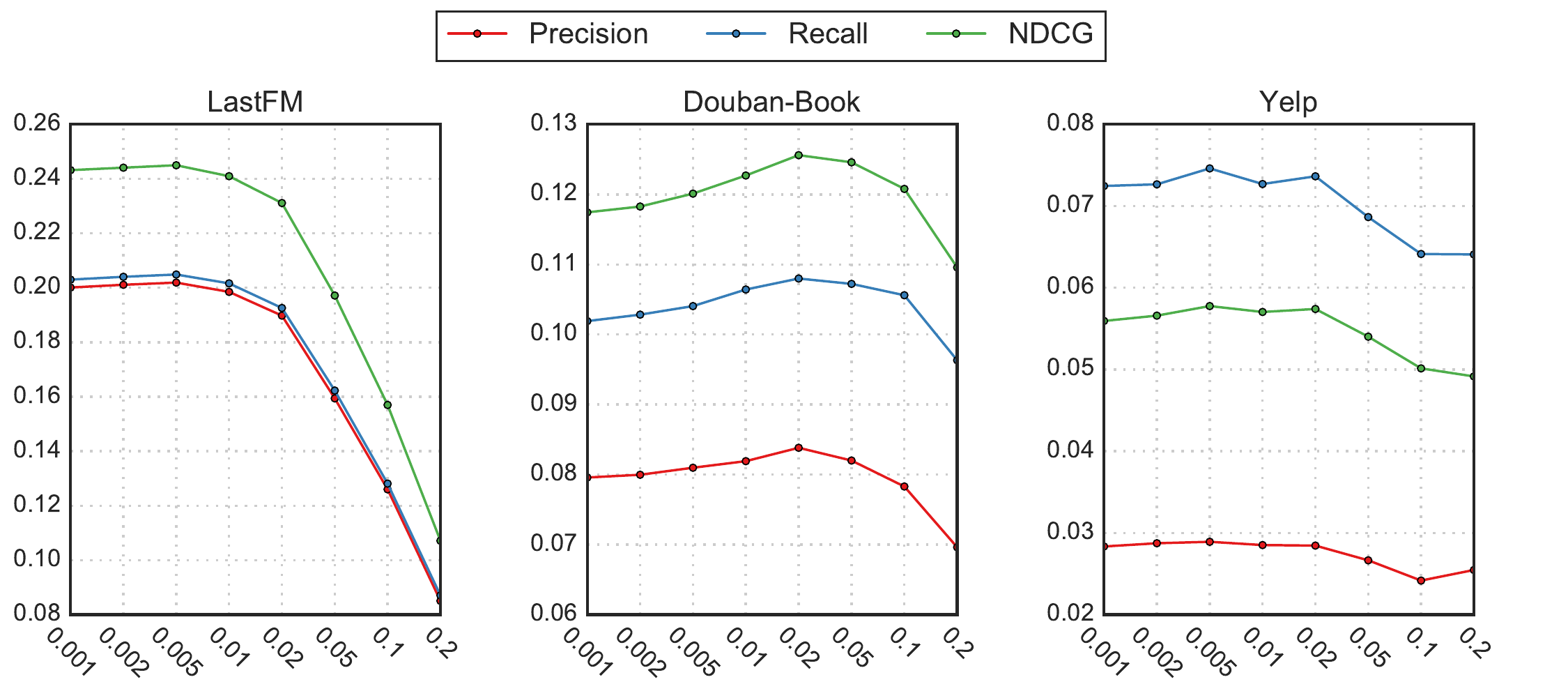}
	\caption{Sensitivity analysis of $\beta$.}
	\label{figure.4}
\end{figure}

\begin{figure}[t]
	\centering
	\includegraphics[width=.5\textwidth]{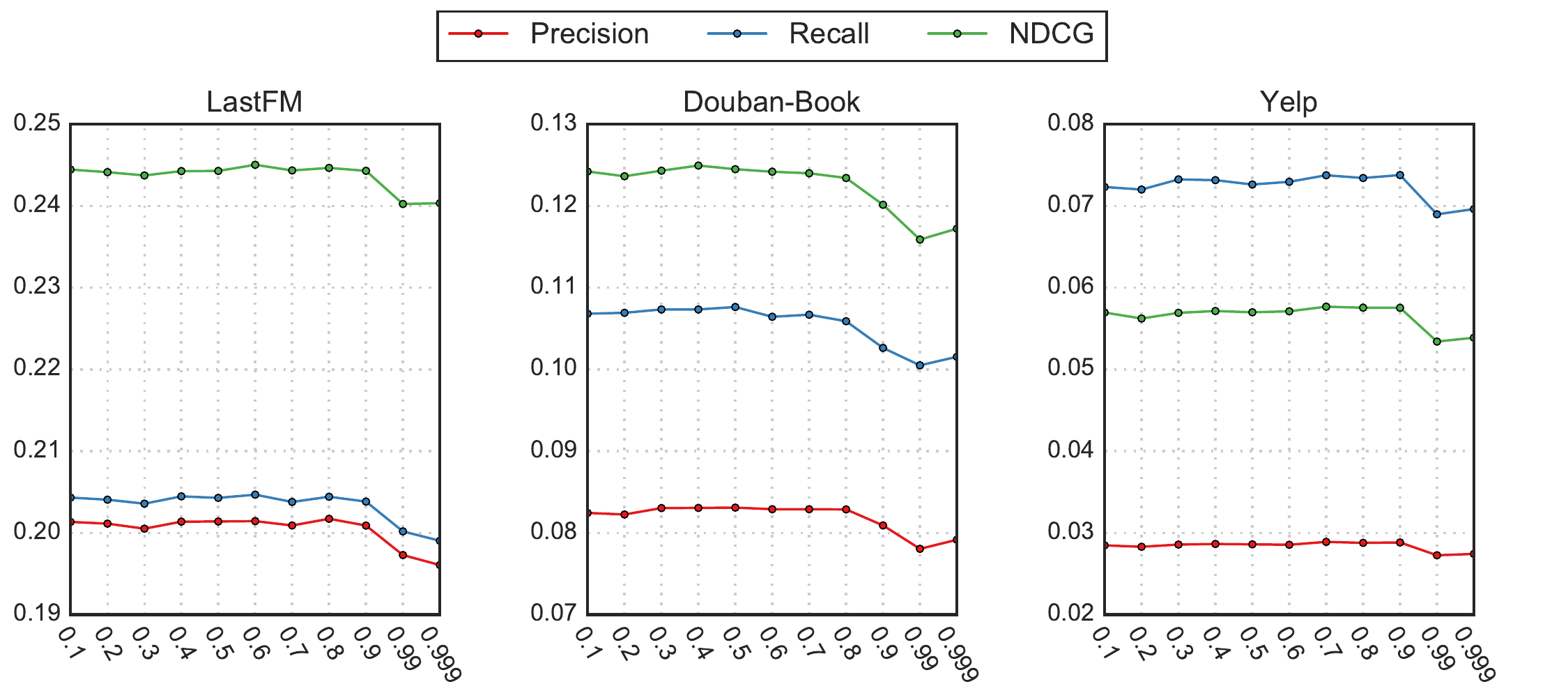}
	\caption{Influence of the edge dropout rate.}
	\label{figure.6}
	\vspace{-10pt}
\end{figure}

\subsection{Parameter Sensitivity Analysis}
There are three important hyper-parameters used in SEPT: $\beta$ for controlling the magnitude of self-supervised tri-training, $K$ - the number of used positive examples and $\rho$ - the edge dropout rate of $\tilde{\mathcal{G}}$. We choose some representative values for them to investigate the parameter sensitivity of SEPT. The results are presented in Fig. \ref{figure.4} - \ref{figure.6}. When investigating the influence of $\beta$, we fix $K=10$ and $\rho=0.3$. For the influence of $K$ in Fig. \ref{figure.5}, we fix $\beta=0.005$ on LastFM and Yelp, $\beta=0.02$ on Douban-Book, and $\rho=0.3$. Finally, for the effect of $\rho$ in Fig. \ref{figure.6}, the setting of $\beta$ is as the same as the last case, and $K=10$. A two-layer setting is used in this case.\par
As can be observed from Fig. \ref{figure.4}, SEPT is sensitive to $\beta$. On different datasets, we need to choose different values of $\beta$ for the best performance. Generally, a small value of $\beta$ can lead to a desirable performance, and a large value of $\beta$ results in a huge performance drop. Figure \ref{figure.5} has been interpreted in Section 4.3. According to Fig. \ref{figure.6}, we observe that SEPT is not sensitive to the edge dropout rate. Even a large value of $\rho$ (e.g., 0.8) can create informative self-supervision signals, which is a good property for the possible wide use of SEPT. When the perturbed graph is highly sparse, it cannot provide useful information for self-supervised learning.

\section{Conclusion and Future work}
The self-supervised graph contrastive learning, which is widely used in the field of graph representation learning, recently has been transplanted to recommendation for improving the recommendation performance. However, most SSL-based methods only exploit self-supervision signals through the self-discrimination, and SSL cannot fully exert itself in the scenario of recommendation to leverage the widely observed homophily. To address this issue, in this paper, we propose a socially-aware self-supervised tri-training framework named SEPT to improve recommendation by discovering self-supervision signals from two complementary views of the raw data. Under the self-supervised tri-training scheme, the neighbor-discrimination based contrastive learning method is developed to refine user representations with pseudo-labels from the neighbors. Extensive experiments demonstrate the effectiveness of SEPT,  and a thorough ablation study is conducted to verify the rationale of the self-supervised tri-training. \par
In this paper, only the self-supervision signals from users are exploited. However, items can also analogously provide informative pseudo-labels for self-supervision. This can be implemented by leveraging the multimodality of items. We leave it as our future work. We also believe that the idea of self-supervised multi-view co-training can be generalized to more scenarios beyond recommendation.
\section*{Acknowledgment}
This work was supported by ARC Discovery Project (Grant No. DP190101985) and ARC Training Centre for Information Resilience (Grant No. IC200100022).

\bibliographystyle{ACM-Reference-Format}
\bibliography{refs}

\end{document}